\documentclass[prc,twocolumn,superscriptaddress,showkeys,showpacs]{revtex4-1}
\pdfoutput=1
\usepackage{graphicx}
\usepackage{dcolumn}
\usepackage{longtable}
\usepackage{amsmath}
\usepackage{amssymb}
\usepackage[pdfstartview=FitH,
CJKbookmarks=true,
bookmarksnumbered=true,
bookmarksopen=true,
colorlinks,
pdfborder=001,
linkcolor=blue,
anchorcolor=blue,
citecolor=blue
]{hyperref}
\usepackage{url}
\begin{document}
\title{Axially Symmetric Quadrupole-Octupole Model incorporating Sextic Potential}
\author{M. Chabab }
\email{mchabab@uca.ac.ma (corresponding author)}
\affiliation{High Energy Physics and Astrophysics Laboratory, Physics Department, Faculty of Sciences Semlalia, Cadi Ayyad University P.O.B. 2390, Marrakesh 40000, Morocco.}
\author{A. El Batoul}
\email{abdelwahed.elbatoul@uca.ac.ma}
\affiliation{High Energy Physics and Astrophysics Laboratory, Physics Department, Faculty of Sciences Semlalia, Cadi Ayyad University P.O.B. 2390, Marrakesh 40000, Morocco.}
\author{L. El Ouaourti}
\email{loubnaouaourti@gmail.com}
\affiliation{High Energy Physics and Astrophysics Laboratory, Physics Department, Faculty of Sciences Semlalia, Cadi Ayyad University P.O.B. 2390, Marrakesh 40000, Morocco.}
\begin{abstract}
We present an extended application of the analytic quadrupole octupole axially symmetric model, originally employed to study the octupole deformation and vibrations in light actinides using an infinite well potential (IW). In this work, we extend the model's applicability to a broader range of nuclei exhibiting octupole deformation by incorporating a sextic potential instead of the Davidson potential.Similarly  to conventional models, such as AQOA-IW (for infinite square potential) and AQOA-D (for the Davidson potential), our proposed model is referred to as AQOA-S.
By employing the sextic potential, phenomenologically represented as $v(\tilde\beta) = a_1\tilde \beta^2+a_2\tilde \beta^4+a_3\tilde \beta^6$,  we can derive analytical expressions for the energy spectra and transition rates (B(E1), B(E2), B(E3)). The energy spectra of the model are essentially governed by two critical parameters: $\phi_0$, indicating the balance between octupole and quadrupole strain, and $\alpha$, a key factor in adjusting the shape and behavior of the spectra through the sextic potential. In terms of applications, the study encompasses five isotopes, namely $^{222-226}$Ra and $^{224,226}$Th. Significantly, our model demonstrates remarkable agreement with the corresponding experimental data, particularly for the recently determined B(EL) transition rates of $^{224}$Ra, surpassing the performance of the model that employs the Davidson potential. The stability of the octupole deformation in $^{224}$Ra adds particular significance to these findings.  
\end{abstract}
\pacs{21.60.Ev, 21.60.Fw, 21.10.Re, 23.20.Js} 
\maketitle
\section{Introduction}
\par The field of nuclear physics is devoted to studying the properties and behavior of atomic nuclei. Among the many areas of interest within nuclear physics, the deformation of atomic nuclei has been a subject of significant focus. Deformation refers to the non-spherical shape of atomic nuclei, which can be characterized by quadrupole and octupole moments. Understanding this deformation is important, as it can have a major impact on the structure and dynamics of the nucleus. In recent years, there has been particular interest in the interplay between the quadrupole\cite{b1} and octupole \cite{b2,b3,b4} deformations in atomic nuclei\cite{b5,b6}. This phenomenon arises due to the asymmetry in the distribution of electric charge within the nucleus, which gives rise to a deformation that can affect nuclear properties such as energy levels and electromagnetic moments.
\par A variety of models have been proposed to explore the collective motion of nuclei with quadrupole and octupole deformations, including the analytic quadrupole-octupole axially symmetric model with an infinite well \cite{b5} and Davidson \cite{b6} potentials prototypes. However, these models have limitations and may not always provide the desired outcomes, requiring alternative potentials to be used. Reflection-asymmetric nuclear shapes have been studied historically in various papers, exploring different theoretical approaches such as the Bohr geometrical approach and the extended interacting boson model proposed by Engel and Iachello \cite{b7} in 1985.  These models  have  been thoroughly applied in recent works \cite{b8,b9,b10, b11,b12}. Shneidman et al. \cite{b13} have also proposed an alternative approach based on a-cluster configurations. While the geometrical approach has been utilized in several theoretical studies \cite{b2,b4,b14,b15,b16,b17} over the past 50 years to investigate octupole vibrations around stable quadrupole deformation, most of these studies are limited to axial symmetry.

\par A distinguishing characteristic of octupole deformation is the presence of a negative-parity band, with energy levels denoted by $L^{\pi}=1^{-},3^{-},5^{-},...$, located in close proximity to the ground-state band. Together, these bands form a single band with alternating parity denoted by $L^{\pi}=0^{+},1^{-},2^+,3^{-},4^+,...$. The negative-parity band, which is systematically higher than the ground-state band, is an indication of the presence of octupole vibrations within the nucleus. 
%---------------------------------------------------------------------------------------------------
It is well known that  the specific physical characteristics of systems displaying reflection asymmetry are tied to the violation of $\mathcal{R}$-symmetry  and  $\mathcal{P}$-symmetry. The systems can still exhibit invariance with respect to the product operator $\mathcal{PR}^{-1}$  even though these symmetries are  independently  violated \cite{b1}. Consequently, the system's spectrum is marked by the presence of energy bands wherein parity alternates in conjunction with angular momentum. Morever, the overall structure of these alternating parity bands can be displayed within a larger framework. In the low-energy realm, the system is characterized by oscillations of the octupole shape between opposing orientations, termed as the soft octupole mode, along with simultaneous rotations of the entire quadrupole-octupole structure. The shift in parity results from tunneling between the two reflection asymmetric shape orientations, separated by an angular momentum-dependent potential barrier \cite{b17a}. As the angular momentum increases, the energy barrier escalates, suppressing both the tunneling effect and shape oscillations \cite{b17b}. This process culminates in the establishment of a stable quadrupole-octupole shape. At higher angular momenta, the parity effect diminishes, enabling properties of collective motion to be linked with the rotation of a stable quadrupole-octupole shape. Similar scenarios, particularly  the parity inversion mechanism proposed by Jolos et al., have been previously elucidated \cite{b17c}.
%---------------------------------------------------------------------------------------------------

\par An alternative potential that has found widespread use in the collective Bohr-Mottelson Hamiltonian is the sextic oscillator potential \cite{b18}, which will be discussed in this paper. The sextic oscillator is an anharmonic oscillator potential that has been found to be a useful approximation for the nuclear potential energy surface for nuclei with axially symmetric shapes. The use of  the sextic oscillator potential in collective Bohr-Mottelson models has been tested in several studies \cite{b19,b20,b21,b22,b23,b24,b25,b26,b27,b28,b29,b30,b31,b32,b33}. These studies have shown that the potential of the sextic oscillator can accurately describe certain nuclear phenomena, especially the coexistence of shapes, which is a difficult phenomenon to observe and study. The coexistence of shapes refers to the existence of two or more different nuclear shapes at the same energy level. This phenomenon has been observed in several atomic nuclei and is an important research topic in nuclear physics. The use of the sextic oscillator potential in the collective Bohr-Mottelson Hamiltonian has enabled the study of this complex phenomenon and has provided a better understanding of the behavior of atomic nuclei. However, it is important to note that while the sextic oscillator potential has been successful in describing certain nuclear phenomena, it may not be suitable for all nuclear systems. Further research is necessary to determine the limits and applicability of this potential in describing the behavior of atomic nuclei.
\par The paper introduces a new version  of the analytical quadrupole octupole axially symmetric (AQOA) model, dubbed AQOA-S, featuring the inclusion of a sextic potential. Its primary objective is to provide a comprehensive description of actinides positioned at the juncture between octupole deformation regions and octupole vibrations.

The model assumes equal importance of quadrupole and octupole deformations, with their relative presence determined by a single parameter, $\phi_0$. Axial symmetry is assumed for simplicity, and the separation of variables is similar to the X(5) model \cite{b34}, which describes the first-order shape phase transition between spherical and quadrupole-deformed shapes. It is important to underscore that the assumptions invoked in our constructed model are based on prior investigation carried out by Bonatsos et al. \cite{b6}. However, an alternative approach to the problem of  phase transition in the octupole mode has been discussed in \cite{b35,b36}, using a new parametrization of the quadrupole and octupole degrees of freedom, taking the intrinsic frame of reference as the principal axes of the overall tensor of inertia resulting from the combined quadrupole and octupole deformation. While comparing AQOA models to AQOA-S proposed approach, three main differences between these models are identified, namely: the analytic nature of the AQOA models, the treatment of quadrupole and octupole degrees of freedom, and the symmetry axes of the deformations.
\par The current study follows a specific plan: it begins with an introduction, followed by the presentation of  the theoretical background of  AQOA-S model in section II. The associated numerical results and discussion are covered in section III. Section IV is devoted to our conclusion.

\section{Theoretical Background of the Model}
When it comes to modeling axially symmetric deformations of quadrupole ($\beta_{2}$) and octupole ($\beta_{3}$) in the nucleus, one of the most famous collective Hamiltonians used is the following\cite{b37,b38,b6}:
\begin{multline}\label{Eq1}
	H = -\sum_{\lambda=2,3} {\hbar^2 \over 2 B_\lambda} {1\over \beta_\lambda^3} 
	{\partial \over \partial \beta_\lambda} \beta_\lambda^3 {\partial \over 
		\partial \beta_\lambda} + {\hbar^2 \hat {L^2} \over 6(B_2 \beta_2^2 + 
		2 B_3 \beta_3^2) } \\ + V(\beta_2,\beta_3)  
\end{multline}
where $\beta_2$ and $\beta_3$ are the quadrupole and octupole deformations, $B_2$, $B_3$ are the mass parameters, 
and $\hat L$ is the angular momentum operator in the intrinsic frame, taken along the principal axes of inertia. 
The solutions of the Schrödinger equation have the following separate form \cite{b37} :
\begin{equation}
	\Phi^{\pm}_L(\beta_2,\beta_3,\theta) = (\beta_2 \beta_3)^{-3/2} 
	\Psi^{\pm}_L(\beta_2,\beta_3) \vert LM_{L}K_{L},\pm\rangle,
	\label{Eq2}
\end{equation}
in the case of an axially symmetric nucleus where the collective rotation is perpendicular to the intrinsic symmetry $z'$-axis; $K$=0, the function $\vert LM_{L}K_{L},\pm\rangle$ is expressed as follows\cite{b1}: 
\begin{equation}
	\vert LM_{L}0, \pm\rangle = \sqrt{ 2L+1 \over 32 \pi^2} (1 \pm (-1)^L ) {\cal D}^L_{0,M_{L}}(\theta), 
	\label{Eq3}
\end{equation}  
$\theta$ are the three Euler angles, ${\cal D}_{L}(\theta)$ is the Wigner function of rotation. The positive and negative signs correspond respectively to symmetric states (L even) and antisymmetric states (L odd).
By making the following changes\cite{b37,b38}:
\begin{multline}\label{Eq4}
	\tilde \beta_2 = \beta_2 \sqrt{B_2\over B}, \quad \tilde \beta_3 = \beta_3 
	\sqrt{B_3\over B}, \quad B= {B_2+B_3 \over 2},
\end{multline}
and introducing polar coordinates\cite{b37,b38,b6}: 
\begin{equation}\label{Eq5}
	\tilde \beta_2 = \tilde \beta \cos \phi, \quad \tilde \beta_3 = \tilde \beta
	\sin \phi, \quad \tilde \beta = \sqrt{\tilde \beta_2^2 + \tilde \beta_3^2}, \\
\end{equation} 
where $ \tilde \beta \geq 0$  is the new radial coordinate and $\phi\in[-\pi/2,\pi/2]$, the collective Schrödinger equation, considering the reduced energies $\epsilon_L=(2B/\hbar^2)E$ and reduced potentials $v=(2B/\hbar^2) V$\cite{b39,b40}, and accounting for the three degrees of freedom, can be written as follows:
\begin{multline}\label{Eq6}
	\left[ -{\partial^2 \over \partial \tilde \beta^2} -{1\over \tilde \beta} 
	{\partial \over \partial \tilde \beta} +{L(L+1) \over 3 \tilde \beta^2 
		(1+\sin^2\phi) } -{1\over \tilde \beta^2} {\partial^2 \over \partial \phi^2} 
	+ v(\tilde \beta,\phi) \right. \\ \left.+ {3\over  \tilde \beta^2 \sin^2 2\phi}-\epsilon_L
	\right] \Psi_L^{\pm}(\tilde \beta,\phi) =0.
\end{multline}
It is apparent that quadrupole deformation alone corresponds to $\phi=0$, whereas octupole deformation alone corresponds to $\phi=\pm\pi/2$. Notably, the transformation described in Eq.(\ref{Eq5}) allows for both positive and negative values of $\beta_3$, while $\beta_2$ is restricted to positive values only.
The potential model employed in this study is a generalized phenomenological model that incorporates a reduced potential, enabling the separation of variables. The specific form of the potential is given by:
\begin{equation}
v(\tilde \beta,\phi) = v(\tilde\beta) + \omega(\tilde \phi^\pm),
\label{Eq7}
\end{equation}
where $\omega(\tilde{\phi}^\pm)$ represents two harmonic oscillators with different minima $\pm \phi_0$. The specific form of $\omega(\tilde{\phi}^\pm)$ is:
\begin{equation}
	\omega (\tilde \phi^\pm )= {1\over 2} c (\phi \mp \phi_0)^2 = {1\over 2} c 
	(\tilde \phi^\pm)^2,
	\label{Eq8}
\end{equation}
where $\tilde{\phi}^\pm = \phi \mp \phi_0$. Additionally, the term involving $\tilde{\beta}$ in the potential is modeled as a sextic potential \cite{b18}:
\begin{equation}
	v(\tilde\beta)=\left(b_{s}^{2}-4 a_{s} c_{s}\right) \tilde{\beta}^2+2 a_{s} b_{s} \tilde{\beta}^4+a_{s}^2 \tilde{\beta}^6.
	\label{Eq9}
\end{equation}
By employing the indicated potential form from Eq. (\ref{Eq7}), the collective Schrödinger equation (\ref{Eq6}) can  approximately be decomposed into two differential equations :
\begin{multline}
	\left[ -{\partial^2 \over \partial \tilde \beta^2} - {1\over \tilde \beta} 
	{\partial \over \partial \tilde \beta} +{1\over \tilde\beta^2} \left(
	{L(L+1) \over 3 (1+\sin^2\phi_0)} + {3\over \sin^2 2\phi_0}\right) \right. \\ \left.
	+v(\tilde \beta) -\epsilon_{\tilde\beta} \right] F (\tilde \beta) 
	=0, \qquad \qquad 
	\label{Eq10} 
\end{multline}
and 
\begin{equation}
	\left[ -{1\over \langle \tilde \beta^2 \rangle}{\partial^2 \over 
		\partial(\tilde \phi^\pm)^2} + \omega (\tilde \phi^\pm ) - \epsilon_{\phi} \right]
	\chi(\tilde \phi^\pm ) =0, 
	\label{Eq11}
\end{equation} 
with $\epsilon_L= \epsilon_{\tilde \beta} +\epsilon_{\phi}$ and $\Psi_L^{\pm}(\tilde \beta,\phi) = N_{\tilde \beta} F(\tilde \beta) N_\phi^{\;\pm} [\chi(\tilde \phi^+) \pm \chi(\tilde \phi^-)]/\sqrt{2}$.
The constants $N_{\tilde \beta}$ and $N_\phi^{\pm}$ are determined from the normalisation of  the total wave function,
while $\langle \tilde \beta^2\rangle$ is the average of 
$\tilde \beta^2$ over $F(\tilde \beta)$ with the integration measure $\tilde \beta d\tilde \beta$.

\subsection{Analytical determination of energy spectra}
\subsubsection{The $\tilde \beta$ equation and its solutions}
It is appropriate to write equation (\ref{Eq10}) in a Schrödinger form. This is achieved by changing the wave function with $F(\tilde{\beta})=\tilde{\beta}^{-1/2}
\xi(\tilde{\beta})$ :
\begin{align}
	\left[-\frac{d^2}{d \tilde{\beta}^2}
	+\frac{1}{\tilde{\beta}^2}\frac{L(L+1)}{3\left(1+\sin ^2 \phi_0\right)}+\frac{3}{\tilde{\beta}^2 \sin ^2 2 \phi_{0}}
	-\frac{1}{4 \tilde{\beta}^2} \right. \nonumber\\ \left.
	+v(\tilde{\beta})\right] \xi(\tilde{\beta})= \epsilon_{\tilde \beta}\; \xi(\tilde{\beta}).
	\label{Eq12}
\end{align}
Furthermore, the earlier differential equation (\ref{Eq12}) can be restructured to align with the quasi-exact characteristics of the sextic oscillator potential featuring a centrifugal barrier, as elucidated in Ref\cite{b18}, through the application of the relationship :
\begin{equation}
\quad \frac{L(L+1)}{3}=\left(2 s-\frac{1}{2}\right)\left(2 s-\frac{3}{2}\right),
%\Rightarrow s=\frac{L}{4}+\frac{3}{4},  $$  and $$ \quad\quad c_{s}=s+\frac{1}{2}+M \Rightarrow c_{s}=\frac{L}{4}+\frac{5}{4}+M.
	\label{Eq13}
\end{equation}
from this relation we can derive the expression of $s$ as a function of $L$ as \cite{b25}:
\begin{equation}
   s(L)=\frac{1}{2}\Bigg[1+\sqrt{\frac{L(L+1)}{3}+\frac{1}{4}}\Bigg].
	\label{Eq13a}
\end{equation}
The collective Schrödinger equation (\ref{Eq12}) under the sextic potential can be significantly simplified by reducing the number of parameters through the variable change of 	$\tilde{\beta}=ya_{s}^{-1/4}$ and adopting the notation $\alpha=b_{s}/\sqrt{a_{s}}$, $\epsilon_{y}= \epsilon_{\tilde \beta}/\sqrt{a_{s}}$ and $\xi(\tilde \beta)=\eta(y)$ :
\begin{align}
	\Big[-\frac{d^2}{d y^2}+\frac{1}{y^2}\frac{L(L+1)}{3\left(1+\sin ^2 \phi_0\right)}+
	\frac{1}{y^2}\frac{3}{ \sin ^2 2 \phi_0}-\frac{1}{4 y^2}\nonumber\\ 
	+v(y)\Big] \eta(y)= \epsilon_{y}\eta(y), 
		\label{Eq14}
\end{align}
where
\begin{equation}
	v(y)=\left(\alpha^2-4  c_{s}\right) y^2+2  \alpha y^4+ y^6.
		\label{Eq15}
\end{equation}
In order to ensure that the energy potential remains invariant across all band states, it is necessary to meet the following condition\cite{b25}:
\begin{equation}
 c_s=s+\frac{1}{2}+M=const. \quad \quad M = 0, 1, 2, ...
	\label{Eq15a}
\end{equation}
It is obvious that the sextic potential also exhibits a dependence on the quantum angular momentum $L$ through its association with the parameter $c_s$. Given that $M$ is an integer, and considering the formula for $s(L)$ provided in Eq. (\ref{Eq13a}), it turn out that this condition cannot be satisfied for all values of the quantum angular momentum $L$ due to $s$ potentially taking on irrational values. Consequently, it is necessary to approximate $s(L)$ as suggested in Ref. \cite{b25} by using :
\begin{equation}
    s'(L)=\frac{L}{4}+\frac{3}{4}.
	\label{Eq15b}
\end{equation}
It is worth noting that this approximation can be justified by using the same approach employed in \cite{b25}. In fact, it suffices to calculate the expressions of $s(L)$ and $s'(L)$ for various states by evaluating $L$. Table (\ref{tab00}) below provides these values for reference.
\begin{table}[htbp]
\caption{A comparison of $s(L)$ and $s^{\prime}(L)$, given by equations (\ref{Eq13a}) and (\ref{Eq15b}) for even $L \leq 10^+$ and odd $L \leq 11^{-}$.}
	\begin{center}
		$
		\begin{array}{|l||cccccc|}
			\hline L^{\pi} & 0^{+} & 2^{+} & 4^{+} & 6^{+} & 8^{+} & 10^{+} \\
			\hline\hline s(L) & 0.75 & 1.25 & 1.81 & 2.39 & 2.96 & 3.54 \\
			s^{\prime}(L) & 0.75 & 1.25 & 1.75 & 2.25 & 2.75 & 3.25 \\
			\hline
		\end{array}
		$
		\qquad \qquad
		$
		\begin{array}{|l||cccccc|}
			\hline L^{\pi} & 1^{-} & 3^{-} & 5^{-} & 7^{-} & 9^{-} & 11^{-} \\
			\hline\hline s(L) & 0.98 & 1.53 & 2.10 & 2.67  & 3.25 & 3.83 \\
			s^{\prime}(L) & 1.00 & 1.50 & 2.00 & 2.50 & 3.00 & 3.50 \\
			\hline
		\end{array}
		$
	\end{center}
\label{tab00}
\end{table}

As a result, the condition expressed in Eq. (\ref{Eq15a}) is now fulfilled and can be expressed as:
\begin{equation}
	c_{s}=s'(L)+\frac{1}{2}+M=\frac{L}{4}+\frac{5}{4}+M = const. \quad \quad M = 0, 1, 2, ...
	\label{Eq16}
\end{equation}
To address the sextic potential  for levels with both negative and positive parity, we examine four distinct constants that correspond to four sets of states:
\begin{itemize}
\item \textbf{For L even} :
\begin{align}
		\label{Eq17}
& (M, L): (K, 0), (K-1, 4), (K-2, 8),\ \cdots ´\Rightarrow c_{0}^{K}=K+\frac{5}{4}\\
& (M, L): (K, 2), (K-1, 6), (K-2, 10),\ \cdots \Rightarrow c_{2}^{K}=K+\frac{7}{4}
	\label{Eq18}
\end{align}
\item  
\textbf{For L odd} :
\begin{align}
		\label{Eq19}
&(M, L): (K, 1), (K-1, 5), (K-2, 9),\ \cdots\Rightarrow c_{1}^{K}=K+\frac{3}{2}\\
&(M, L): (K, 3), (K-1, 7), (K-2, 11),\ \cdots \Rightarrow c_{3}^{K}=K+2	
	\label{Eq20}
\end{align}
\end{itemize}
where: 
$K=M_{max}$ + $L_{min}/4$, once $M$ increases by one unit $L$ must decrease by four units to keep $K$ constant, the four formula of $c^{K}$ can be summed up by the expression :
\begin{equation}
	c_{m}^{K}= c_{0}^{K}+\frac{1}{4}m=K+\frac{5}{4}+\frac{1}{4} m ,\ \  m = 0, 1, 2, 3.,
		\label{Eq21} 
\end{equation}
The condition of a constant potential is indeed fulfilled for four distinct sets of states, each corresponding to slightly different potentials. To further enhance this understanding, the general potential can be considered in the following form:
\begin{equation}
 v_{m}^{K}(y)=\left(\alpha^2-4  c_{m}^{K}\right) y^2+2  \alpha y^4+ y^6+u_{m}^{K}(\alpha).
 	\label{Eq22}
\end{equation}
The constants $u_m^K(\alpha)$ will be determined in such a way that the minima ($y_{0,m}^{K} > 0$) of the four potentials have identical energy. 
The extremal points can be obtained from the first derivative of the potential. Therefore:
\begin{equation}
\left(y_{0,m}^{K}\right)^2 =\frac{1}{3 }\left[-2 \alpha + \sqrt{\alpha^2+ 12c_{m}^{K}}\right].
	\label{Eq23}
\end{equation}
By selecting $u_0^K = 0$, the remaining constants can be determined as follows:
\begin{align}
	u_{i}^{K}=&
	\left(\alpha^2-4  c^{K}_{0}\right) (y^{K}_{0,0})^2- \left(\alpha^2-4  c_{i}^{K}\right) (y_{0,i}^{K})^2 \nonumber\\+&
	2 \alpha \left((y^{K}_{0,0})^4-(y_{0,i}^{K})^4 \right) + (y^{K}_{0,0})^6-(y_{0,i}^{K})^6, \quad i= 1, 2, 3  
		\label{Eq24} 
\end{align}
The previous details were established to derive a solvable sextic equation for the axially symmetric nucleus with quadrupole-octupole deformation. Now, we shift our attention to solving Eq. (\ref{Eq14}). For this purpose, the following ansatz is considered \cite{b18} :
\begin{equation}
	\eta^{\;(M)}(y) \sim P^{(M)}_n\left(y^2\right) y^{2 s'(L)-\frac{1}{2}} \mathrm{e}^{-\frac{y^4}{4}-\frac{\alpha y^2}{2}}.
		\label{Eq25}
\end{equation}
By replacing this wave function  in equation (\ref{Eq14}), we get a quasi-exactly solvable equation :
\begin{align}
	\Bigg[-\left(\frac{d^2}{d y^2}+\frac{4s'-1}{y}\frac{d}{d y}\right)&+2\alpha y \frac{d}{d y}+2y^{2}\left(y \frac{d}{d y} -2M \right)\nonumber\\ &\Bigg]
	P^{(M)}_n\left(y^2\right) = \lambda  P^{(M)}_n\left(y^2\right),\qquad\qquad\qquad\qquad
	\label{Eq26}
\end{align}
where $P^{(M)}_n\left(y^2\right)$ are polynomials in $y^2$ of order $M$. To obtain the eigenvalues $\lambda$ for each $M$, we can employ the same analytical procedure outlined in Ref. \cite{b21}. For a given value of $M$, there are $M+1$ solutions distinguished by the $\beta$ vibrational quantum number $n=n_{\beta}$. The lowest eigenvalue corresponds to $n_{\beta}=0$, while the highest corresponds to $n_{\beta}=M+1$. The eigenvalues $\lambda$ also depend on $L$ through the parameter $s'$, and it should be noted that $L$ and $M$ are interconnected through the condition (\ref{Eq16}), with the specific relationship being determined by the value of $K$. Hence, in the subsequent analysis, we will replace the $M$ indexing of $\lambda$  with $K$. After performing the necessary algebraic manipulations leading to Eq. (\ref{Eq26}) and considering the aforementioned considerations, the eigenvalues $\lambda$ can be alternatively expressed as:
\begin{equation}
\lambda=\lambda_{n_{\beta},L}^{M}= \epsilon_{y} - u^{K}_{m}(\alpha)  - 4\alpha s - 
\frac{W_{L}(\phi_0)}{\langle y^2\rangle_{M L}},
\label{Eq27}
\end{equation}
with 
\begin{equation}
W_{L}(\phi_0)=
\frac{3}{ \sin ^2 2 \phi_{0}}+\frac{L(L+3)-2}{4\left(1+\sin ^2 \phi_0\right)}-\frac{1}{4}(L+1)^{2}.
\label{Eq28}
\end{equation}
The $\beta$ part of the total energy is then :
\begin{equation}
E_{n_{\beta},L}=\frac{\hbar^{2}\sqrt{a_{s}}}{2B}\left[ \lambda_{n_{\beta},L}^{M} +\alpha(L+3)+ u^{K}_{m}(\alpha) + \frac{W_{L}(\phi_0)}{\langle y^2\rangle_{M L}}
	\right].
	\label{Eq29}
\end{equation}
Here, $\langle y^2\rangle_{M L}$ represents the average value of $y^2$ over the wave functions $\eta^{(M)}(y)$ given by Eq.(\ref{Eq25}), with the polynomials $P^{(M)}_n\left(y^2\right)$ organized in Table (\ref{tab0}) based on their corresponding values of $M$. The integration measure is denoted as $dy$. 
It should be noted that the coefficients $C_{Mn}$  in the function $P^{(M)}_n$ expressions listed in Table (\ref{tab0}), are determined using the same methodology as in previous studies \cite{b21,b26,b30,b31}.
The eigenvalues $\lambda_{n_{\beta},L}^{M}$ are obtained by diagonalizing the matrix representation of Eq. (\ref{Eq26}). For further elaboration and in-depth insights, we recommend referring to the cited Refs. \cite{b18,b21,b31}. 
\begin{table}[htbp]

	\caption{The polynomial functions  $P^{(M)}_n\left(y^2\right)$ corresponding to each value of $M$.}	
	\begin{center}
		\begin{tabular}{|c|c|}
			\hline
			$M$ & $P_n^{(M)}\left(y^2\right)$ \\
			\hline
			0 & $C_{00}$ \\
			\hline
			1 & $C_{10}+C_{11} y^2$ \\
			\hline
			2 & $C_{20}+C_{21} y^2+C_{22} y^4$ \\
			\hline
			3 & $C_{30}+c_{31} y^2+C_{32} y^4+C_{33} y^6$ \\
			\hline
			4 & $C_{40}+C_{41} y^2+C_{42} y^4+C_{43} y^6+C_{44} y^8$ \\
			\hline
		\end{tabular}
	\end{center}
	\label{tab0}
\end{table}
\subsubsection{The $\phi$ equation and its solutions}
The differential equation (\ref{Eq11}) with the potential $\omega (\tilde \phi^\pm)$ given by Eq. (\ref{Eq8}) is evidently a one-dimensional harmonic oscillator equation. The solution to this equation has been derived and documented in Ref. \cite{b5}. Hence, the formula for the energy eigenvalues can be expressed as follows \cite{b5}:
\begin{equation}
	\epsilon_{\phi} = \sqrt{ 2c\over \langle \tilde \beta^2\rangle } \left(
	n_\phi +{1\over 2} \right), \qquad n_\phi = 0,1,2,\ldots 
	\label{Eq30}
\end{equation}
and eigenfunctions one are\cite{b5} :
\begin{equation}
\chi_{n_\phi}(\tilde \phi^\pm) = N_{n_\phi} H_{n_\phi} (b \tilde \phi^\pm) 
e^{-b^2 (\tilde \phi^\pm)^2 /2} ,
\label{Eq31}
\end{equation}
where 
\begin{equation}
 b=\left( c \langle \tilde \beta^2 
\rangle \over 2 \right)^{1/4}, \quad  N_{n_\phi} = \sqrt{b \over  \sqrt{\pi} 2^{n_\phi} n_\phi!}.
\label{Eq32}
\end{equation}
Taking into account the aforementioned conditions, the total energy of the system can be expressed as: 
\begin{equation}
\textbf{E}_{n_\beta, n_\phi, L}=E_0 + E_{n_\beta,L} + E_{n_\phi}, 
\label{Eq33}
\end{equation}
where
\begin{align}
 &E_0= {\hbar^2 \over 4B}\sqrt{{2 c\over \langle \tilde{\beta}^2 \rangle}},
&E_{n_\phi} = {\hbar^2 \over 2B} \sqrt{{2c\over \langle \tilde{\beta}^2 \rangle}} \times n_\phi,
\label{Eq34}
\end{align}
with $E_{n_\beta,L}$ is given by Eq.(\ref{Eq29}).
In this paper, we focus solely on bands with the quantum number $n_\phi=0$. As a result, this specific quantum number will be excluded from the expressions of the total wave functions. Similarly, the notation $n_\beta$ will be replaced with a simplified notation $n$ in the wave function expressions,
\begin{align}
	\Phi^{\pm}_{n, L, M_{L}}& (\tilde\beta, \phi, \theta)=N_{\tilde \beta}N_\phi^{\pm} (\beta_2 \beta_3)^{-3/2}  F_{n, L}^{(M)}(\tilde \beta)\; \nonumber\\& {[\chi(\tilde \phi^+)\pm \chi(\tilde \phi^-)]\over \sqrt{2}} \sqrt{2L+1\over 32 \pi^2} \times(1\pm (-1)^L) {\cal D}^L_{0,M_{L}}(\theta),
	\label{Eq35}
\end{align} 
with 
\begin{equation}
	F_{n,L}^{(M)}(\tilde\beta)=\tilde{\beta}^{-1/2}P^{(M)}_n\left(\tilde\beta^2\right) \tilde\beta^{2 s'(L)-\frac{1}{2}} \mathrm{e}^{-\frac{a_s\tilde\beta^4}{4}-\frac{b_s\tilde\beta^2}{2}}.
\end{equation}
\subsection{Determination of $B(EL)$ transition rates}
\par A deeper understanding of phenomena in nuclear physics, including alterations in nuclear shape, collective vibrations, and electromagnetic excitations within atomic nuclei, heavily depends on the pivotal role of $B(EL)$ transition rates. These rates serve as quantitative measures, providing valuable information about the strength and probability of specific electromagnetic transitions. Their significance lies in their ability to facilitate the interpretation of experimental data and refine theoretical models in the field of nuclear physics. Through meticulous examination of $B(EL)$ transition rates, we can acquire valuable insights, leading to a more profound comprehension of the intricate dynamics of nuclei and contributing to the advancement of knowledge in fundamental nuclear processes. 
\par In this section, our attention is directed towards the electrical transitions E1 (dipole), E2 (quadrupole), and E3 (octupole). These transitions specifically involve changes in the electric field within atomic nuclei and are of particular interest in our analysis. By studying these electrical transitions, we aim to gain a deeper understanding of their properties and implications for the overall behavior of nuclei.
\par  In case of an axially symmetric nucleus, the electric dipole operator reads \cite{b37},
\begin{equation}
	T^{(E1)}_\mu  = t_1 {B\over \sqrt{B_2 
			B_3}} \tilde\beta^2 {\sin 2\phi \over 2} {\cal D}^{(1)}_{0,\mu}(\theta),
		\label{Eq36}
\end{equation}
while the electric quadrupole and octupole ones are\cite{b1}
\begin{align}
	\label{Eq37}
		&T^{(E2)}_\mu = t_2 \sqrt{B\over B_2} \tilde \beta \cos\phi {\cal D}^{(2)}_{0,\mu}(\theta), \\
		&T^{(E3)}_\mu = t_3 \sqrt{B\over B_3} \tilde \beta \sin\phi {\cal D}^{(3)}_{0,\mu}(\theta), 
	\label{Eq38}
\end{align}
with $t_1$, $t_2={3 Z e\over 4\pi} R^2$ and $t_3={3 Z e\over 4\pi} R^3$ are scale factors, where $R$ is the effective radius of the nucleus, given by \cite{b41} : 
\begin{equation}
	R= r_0 A^{1/3}, \qquad r_0=1.2 \ {\rm fm}, 
	\label{Eq39}
\end{equation}
with $A$ being the mass number of the nucleus. Based on the definition of the coefficients $t_2$ and $t_3$, we can derive the subsequent constraint \cite{b6}:
\begin{equation}
	{t_3\over t_2}= R, 
	\label{Eq40}
\end{equation}
which is particularly useful in practical scenarios when determining the numerical values of $B(EL)$ transition rates for specific atomic nuclei. Additionally, $B(EL)$ transition rates are given by 
\begin{equation}
	B(EL; L_i n_i \to L_f n_f) = { \vert \langle L_f n_f \vert \vert T^{(EL)}
		\vert \vert L_i n_i \rangle \vert^2 \over (2L_i+1)}
	\label{Eq41}
\end{equation}
where the reduced matrix element is obtained through the Wigner-Eckart theorem
\begin{align}
	\langle L_f n_f 
	\vert \vert   T^{EL} \vert \vert L_i n_i \rangle &= { \sqrt{2L_f+1} \over (L_i L L_f \vert \mu_i \mu \mu_f) }\nonumber\\
	&  \times \langle L_f \mu_f n_f \vert T^{(EL)}_\mu \vert L_i \mu_i n_i\rangle 
	\label{Eq42}
\end{align}
The volume element is $d\tau =(\beta_2 \beta_3)^3 d\beta_2  d\beta_3 d^3\theta $, 
using Eqs. (\ref{Eq4}) and (\ref{Eq5}), as well as the relevant Jacobian, one finds that 
$ d\beta_2 d\beta_3 =  {B\over \sqrt{B_2 B_3}} \times \tilde \beta d\tilde \beta d\phi$.
Hence, it is straightforward to compute the matrix elements in Eq. (\ref{Eq35}) for our model as follows:
\begin{align}
\langle L_f n_f \vert \vert T^{(EL)} \vert \vert L_i n_i \rangle = FL \times I_{\theta}^{(EL)} I_{\phi}^{(EL)} I_{\tilde \beta}^{(EL)},
	\label{Eq43}
\end{align}
with
\begin{align}
		\label{Eq44}
&FL=F1 = {1\over 2} t_1 {B \over \sqrt{B_2 B_3}} {e^{-{1\over  b^2}} \sin 2\phi_0 \over \sqrt{1- e^{-2b^2 \phi_0^2}}} \quad \text{for}\ E1,\\
	\label{Eq45}
&FL=F2 = t_2 \times \sqrt{B/B_2} \quad \text{for}\  E2,\\
&FL=F3 = t_3 \times \sqrt{B/B_3} \quad \text{for} \ E3,
	\label{Eq46}
\end{align}
it is worth emphasizing that we have employed a similar methodology to that described in \cite{b6} for the numerical computation of these quantities.
The quantities  $I_{\theta}^{(EL)}$ correspond to the integrals of three Wigner functions over $d^3\theta $, yielding the expression $\sqrt{2L_i+1}\times(L_i L L_f | 0 0 0)$.  On the other hand, $I_{\phi}^{(EL)}$ involve integrals over $\phi$, and have distinct expressions for different possible transitions, as detailed in \cite{b6}:
\begin{align}
	\label{Eq47}
&I_{\phi,S\to S}^{(E2)}= e^{-{1\over 4b^2}} { \cos\phi_0+ e^{-b^2 \phi_0^2} \over 1+ e^{-b^2 \phi_0^2}}, \\
	\label{Eq48}
&I_{\phi,A\to A}^{(E2)}= e^{-{1\over 4b^2}} { \cos\phi_0- e^{-b^2 \phi_0^2} \over 1- e^{-b^2 \phi_0^2}},\\
	\label{Eq49}
&I_{\phi,S\to A}^{(E3)}= {e^{-{1\over 4 b^2}} \frac{\sin\phi_0}{\sqrt{1- e^{-2b^2 \phi_0^2}}}}, \\
&I_{\phi,S\to A}^{(E1)}= {e^{-{1\over b^2}}  \frac{\sin2\phi_0}{2 \sqrt{1- e^{-2b^2 \phi_0^2}}}}.
	\label{Eq50}
\end{align}
The integrals $I_{\tilde \beta}^{(EL)}$  are computed by integrating over $\tilde \beta d\tilde \beta$. The factors 
$(\beta_2 \beta_3)^3$ and  ${B\over \sqrt{B_2 B_3}}$ arise from the volume element and are subsequently simplified by the third term in Eq. (\ref{Eq35}) and $N_{\tilde \beta}^{2}$, respectively. Here, we present the expressions for these integrals:
\begin{align}
	\label{Eq51}
    &I_{\tilde \beta}^{(E3)} =a^{-1 / 4}_{s} \int_0^{\infty} \eta_{n_i, L_i}^{(M)}(y)\; y\; \eta_{n_f, L_f}^{(M)}(y)\; \mathrm{d}y, \\
    \label{Eq52}
    &I_{\tilde \beta}^{(E2)} =a^{-1 / 4}_{s} \int_0^{\infty} \eta_{n_i, L_i}^{(M)}(y)\; y\; \eta_{n_f, L_f}^{(M)}(y)\; \mathrm{d} y,\\
    \label{Eq53}
	&I_{\tilde \beta}^{(E1)}=a^{-1 / 2}_{s} \int_0^{\infty} \eta_{n_i, L_i}^{(M)}(y)\; y^{2}\; \eta_{n_f, L_f}^{(M)}(y)\; \mathrm{d} y.
\end{align}
Lastly, we offer the matrix element expressions for the various possible transitions:
\begin{itemize}
\item[-]  Matrix elements of $T^{(E2)}$ between symmetric states read :
\begin{multline}
	\langle L_f n_f \vert \vert T^{E2} \vert \vert L_i n_i \rangle_{S\to S} = 
	t_2 \sqrt{B\over B_2} e^{-{1\over 4b^2}} { \cos\phi_0+ e^{-b^2 \phi_0^2} \over 1+ e^{-b^2 \phi_0^2}} \\ 
	\sqrt{2L_i+1}(L_i 2 L_f | 0 0 0) \times I_{\tilde \beta}^{(E2)}.\qquad
	\label{Eq54}
\end{multline}
\item[-] Matrix elements of $T^{(E2)}$ between antisymmetric states are
\begin{multline}
	\label{Eq55}
	\langle L_f n_f \vert \vert T^{E2} \vert \vert L_i n_i \rangle_{A\to A} = 
	t_2 \sqrt{B\over B_2} e^{-{1\over 4b^2}} { \cos\phi_0- e^{-b^2 \phi_0^2} \over 1- e^{-b^2 \phi_0^2}} \\
	\sqrt{2L_i+1} (L_i 2 L_f | 0 0 0)\times I_{\tilde \beta}^{(E2)}.\qquad
\end{multline}
\item[-]Matrix elements of $T^{(E3)}$ between a symmetric state and an antisymmetric state have the form :
\begin{multline}
	\label{Eq56}
	\langle L_f n_f \vert \vert T^{E3} \vert \vert L_i n_i \rangle_{S\to A} = 
	t_3 \sqrt{B\over B_3} {e^{-{1\over 4 b^2}}  \frac{\sin\phi_0}{\sqrt{1- e^{-2b^2 \phi_0^2}}}} \\
	\sqrt{2L_i+1}(L_i 3 L_f | 0 0 0)\times I_{\tilde \beta}^{(E3)}.\qquad
\end{multline}
\item[-]  Matrix elements of $T^{(E1)}$ between a symmetric state and an antisymmetric state are 
\begin{multline}
	\label{Eq57}
	\langle L_f n_f \vert \vert T^{E1} \vert \vert L_i n_i \rangle_{S\to A} = 
	{1\over 2} t_1 {B \over \sqrt{B_2 B_3}} {e^{-{1\over  b^2}} \frac{\sin 2\phi_0}{\sqrt{1- e^{-2b^2 \phi_0^2}}}} \\
	\sqrt{2L_i+1}(L_i 1 L_f | 0 0 0)\times I_{\tilde \beta}^{(E1)}.
\end{multline} 
\end{itemize}
\section{Numerical results and discussion}
\par The evaluation of a nuclear structure model often involves assessing how well the calculated spectra align with experimental data, which serves as a measure of the model's quality. A commonly used metric to evaluate the fit's quality is the standard deviation, denoted as $\sigma$. It is calculated as follows: 
\begin{equation}\label{Eq58}
	\sigma=\sqrt{\frac{\sum_{i=1}^{n}\left(E_{i}(\exp )-E_{i}(\mathrm{th})\right)^{2}}{(n-1) E\left(2_{g.s}^{+}\right)^{2}}},
\end{equation}
which quantifies the differences between the calculated and experimental spectra. 
\par If the value of $n_{\phi}$ is zero, the overall energy supplied by the AQOA-S model is influenced by several factors. These factors include the variables $\alpha$, $\phi_0$, $E_0$, as well as $(\hbar^{2}\sqrt{a_{s}})/2B$.
\par As a consequence of the scaling property inherent in the collective model, such as this, numerical applications concerning actual energy spectra are typically carried out using energy ratios. In this way, we have conducted calculations specifically for the spectra of Ra and Th isotopes located at the boundary between the regions characterized by octupole deformation and octupole vibrations, as well as within the aforementioned region.
\par
\begin{figure}[h!]
	\centering
	\includegraphics[width=1\linewidth, height=0.3\textheight]{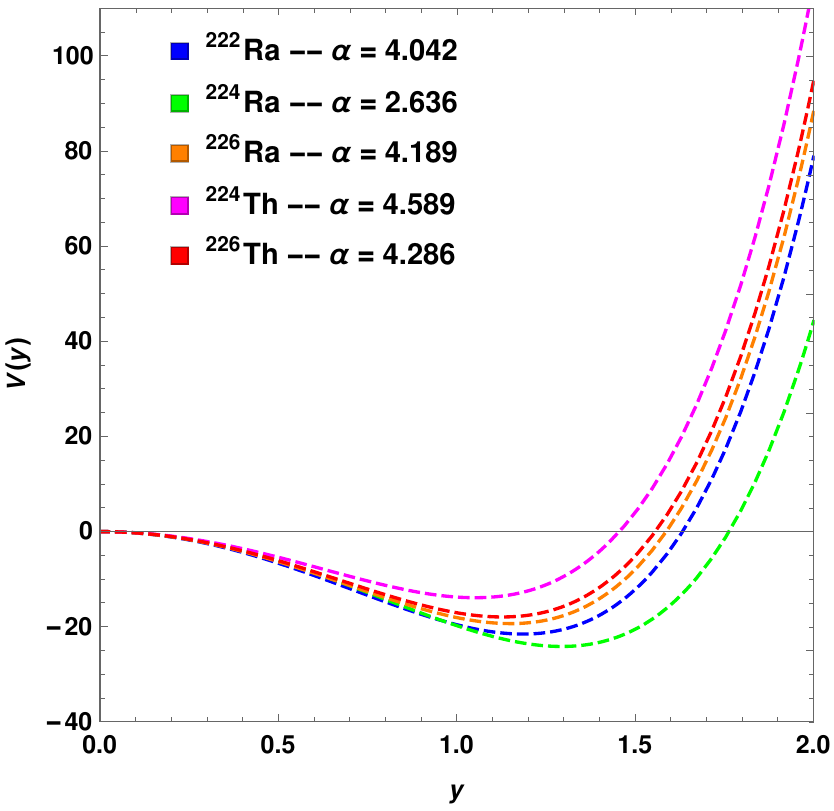}
	\caption[Potential]{The shape evolution of the sextic potential $v(y)$ (Eq. (\ref{Eq15})) are plotted as a function of the scaling quadrupole deformation $y=a_s^{1/4}\beta$ for the isotopes ${ }^{222-226} \mathrm{Ra}$ and ${ }^{224-226} \mathrm{Th}$.}
	\label{Fig1}
\end{figure}

 Table (\ref{Tab1}) displays the obtained values of $\sigma$, as well as the adjusted parameters we used to plot the evolution of the potentials shown in Fig.(\ref{Fig1}), for our model applied to these isotopes. The observed trend in the axially symmetric quadrupole-octupole model, incorporating a sextic potential represented in the said figure (Fig.(\ref{Fig1})), can highlights an interesting relationship between the depth of the minimum potential position and the stability of the nuclei concerning quadrupole/octupole deformation. The deeper the minimum potential position, indicating a lower energy state, the more stable the nuclear core becomes. This increased stability suggests that the forces binding the protons and neutrons within the nucleus are strong enough to withstand deformations, particularly those related to quadrupolar and octupolar shapes. Consequently, nuclei such as $^{224}$Ra,  with deeper potential minima exhibit a greater resilience to shape changes, making them more robust and less prone to structural instabilities.  
\par On the other hand, Table (\ref{Tab2}) summarizes the corresponding spectra that are relevant for these isotopes. Additionally, the predictions of the AQOA-D and AQOA-IW models are included in both tables for the purpose of comparison.
\begin{table*}[htp]
	\caption{Parameters $\phi_0$ and $\alpha$ of the AQOA-S model obtained from rms fits to experimental spectra of ${ }^{222} \mathrm{Ra}$ \cite{b42,b43}, ${ }^{224} \mathrm{Ra}$ \cite{b42,b43}, ${ }^{226} \mathrm{Ra}$ \cite{b42,b43}, $ { }^{224} \mathrm{Th}$ \cite{b44}, and ${ }^{226} \mathrm{Th}$ \cite{b45}. The experimental ratios $R_{4/2}= E(4^{+}_{g.s})/E(2^{+}_{g.s})$ are also displayed. The angular momenta of the highest levels of the ground state, $\beta$ and negative parity bands included in the rms fit are labelled by $L_g$, $L_\beta$ and $L_o$ respectively, while $n$ indicates the number of fitted states. Comparison of the corresponding quality measure $\sigma_{S}$ with the AQOA-D solution \cite{b6} is also shown, as well as the scale factor $\frac{\hbar^{2}\sqrt{a_{s}}}{2B}$.}
	$$\label{Tab1}
	\begin{array}{lcccccccccc}
		\hline \hline \text { Nucleus } & R_{4 / 2} & \phi_0 & \alpha & L_g & L_\beta & L_o & n & \sigma_{S} &  \sigma_{D} & \frac{\hbar^{2}\sqrt{a_{s}}}{2B}[MeV]\\
		\hline
		& &  &  &  &  &  &  & & & \\
		{ }^{222} \mathrm{Ra} & 2.715 & 26.61^{\circ} & 4.042 & 20 & 0 & 19 & 20 & 0.604 & 0.917&0.058 \\
		{ }^{224} \mathrm{Ra} & 2.970 & 38.87^{\circ} & 2.636 & 28 & 0 & 27 & 28 & 0.801 & 1.351&0.064\\
		{ }^{226} \mathrm{Ra} & 3.127 & 15.66^{\circ} & 4.189 &  28 & 0 & 27 & 28 & 1.232 & 1.360 &0.044 \\
		{ }^{224} \mathrm{Th} & 2.896 & 20.00^{\circ} & 4.589 & 18 & & 17 & 17 & 0.691 & 0.843&0.049 \\
		{ }^{226} \mathrm{Th} & 3.136 & 15.53^{\circ} & 4.286 & 20 & 0 & 19 & 20 & 1.059 &0.994&0.051\\
		& &  &  &  &  &  & & & &   \\
		\hline \hline
	\end{array}
	$$
\end{table*}
\begin{table*}
	\caption{Comparison of theoretical predictions of the AQOA-S model, AQOA-D \cite{b6} and the AQOA-IW solution \cite{b5} (for $^{226}$Ra and $^{226}$Th), 
		to experimental energy spectra (normalized to $E(2_1^+)$) of the ground-state, the $\beta$ and the negative parity bands for the considered nuclei.}
	\rotatebox{0}{
		\begin{tabular}{|r || r r r r r r r r r r r r r r r r r|}
			\hline 
			& $^{222}$Ra & $^{222}$Ra & $^{222}$Ra & $^{224}$Ra & $^{224}$Ra & $^{224}$Ra & $^{226}$Ra & $^{226}$Ra &$^{226}$Ra & $^{226}$Ra & $^{224}$Th & $^{224}$Th & $^{224}$Th & $^{226}$Th & $^{226}$Th & $^{226}$Th& $^{226}$Th \\
			$L^{\pi}$ & exp. & S    & D       & exp.       & S  &D & exp.       & S  & D    & IW       & exp.       & S      & D & exp.   & S & D     & IW \\
			
			\hline \hline
			& & &  & & &  & & & &  & &  & & & & &\\ 
			$ 4^+$ & 2.72  &2.64& 3.00& 2.97&2.93& 3.17& 3.13& 3.35  &3.22&  3.09 & 2.90  &3.07 & 3.09& 3.14  & 3.50 &3.22& 3.12 \\ 
			
			$ 6^+$ & 4.95 & 4.65 &5.59 &5.68& 5.57 &6.21& 6.16& 5.93 & 6.45& 5.99 & 5.45 &5.31 &5.90&6.20 &6.19&6.44&6.10 \\
			
			$ 8^+$  & 7.58 & 7.46 &8.49 &8.94&8.42 & 9.87& 9.89& 9.93 & 10.45& 9.56 & 8.50 &8.67 & 9.17&10.00 & 10.43&10.42&9.78\\
			
			$10^+$& 10.55 & 10.14 &11.58 &12.66 &12.35 &13.94&14.19& 13.55& 15.02& 13.71 &  11.97 &11.70 & 12.71& 14.41&14.19&14.97&14.08 \\
			
			$12^+$& 13.82 & 13.88 &14.77 & 16.74&15.83 &18.30& 18.93  & 18.88 &20.00& 18.42 & 15.80  &16.06 &16.42&19.32 &19.80&19.93&18.96 \\
			
			$14^+$ & 17.39 &17.04 &18.04 & 21.17&20.85 & 22.85& 24.06 & 23.24 &25.31& 23.64 & 19.97 &19.68&20.25&24.68 &24.29&25.20&24.38 \\
			
			$16^+$  & 21.21 & 21.58&21.36 & 25.90 &24.83 & 27.55& 29.52  &29.68  &30.84& 29.38 & 24.44 &24.85 &24.16&30.41 &31.06&30.69&30.34 \\
			
			$18^+$  & 25.28&25.10&24.70& 30.92 &30.80 & 32.34 &35.30 & 34.60 &36.54& 35.61 & 29.20 &28.90 &28.13&36.50  &36.05&36.35&36.81 \\
			
			$20^+$ & 29.57& 30.32& 28.07 & 36.22 &35.17 &37.22& 41.38  & 42.02 &42.38& 42.33 &       &       && 42.90 & 43.81&42.14&43.80 \\
			
			$22^+$  & & & &  41.74& 42.00 & 42.15& 47.75 & 47.34 &48.32& 49.54 &       &       &&       &      & &      \\
			
			$24^+$  & & & & 47.48&46.70 & 47.13& 54.44  & 55.60 &54.34& 57.22 &       &       & &      &       & &       \\
			$26^+$  & & & & 53.41 &54.32 &52.14& 61.42& 61.82 &60.43& 65.38 &       &       &  &     &       & &    \\
			$28^+$   & & & & 59.54& 59.31& 57.18 &68.70  & 70.23 &66.57& 74.01 &       &       & &     &       & &       \\
			
			$ 0^+$ & 8.23 & 7.85& 8.06  &10.86&10.99 &10.90&12.19& 10.61&12.21& 11.23  &  & &9.91&11.18& 10.69&11.31& 12.41\\
			&  & &  & & &   &  & & & & & & &  &&&\\
			$ 1^-$ &  2.18 &0.42 &0.35 &2.56& 0.22&0.34&3.75 &0.37 & 0.34& 0.34 &  2.56 & 0.41  & 0.34&  3.19 & 0.37 &0.34&   0.34 \\
			$ 3^-$ &2.85  & 1.69 & 1.90  &3.44& 1.90&1.95&4.75&1.83&1.97& 1.93 &  3.11 & 1.80  & 1.93& 4.26 & 1.88 &1.97&   1.94 \\
			$ 5^-$ & 4.26&3.60  &4.24 &5.13& 4.25&4.60&6.60&4.57 & 4.72 & 4.45 &  4.74 & 4.17  &4.43&  6.24 & 4.79 & 4.72&   4.51 \\
			$ 7^-$ &  6.33  & 5.77&7.01 &7.59&6.96 &7.98&9.26&7.42&8.36& 7.70 &  7.13 & 6.63  & 7.49& 9.11 & 7.79  &8.35&   7.86 \\
			$ 9^-$ &  8.92 & 8.77&10.02 &10.73& 10.68&11.86&12.68&11.70 &12.67& 11.57 & 10.17 & 10.19 &10.91& 12.79 & 12.31 &12.64&  11.86 \\
			$11^-$ & 11.97 & 11.55 & 13.17&14.46& 14.06&16.09&16.74&15.49 &17.47& 16.00 & 13.73 & 13.40 &14.55& 17.15 & 16.32 &17.41&  16.45 \\
			$13^-$ & 15.38& 15.44& 16.40 & 18.68 &18.93 &20.56&21.39  &21.03 &22.63& 20.96 & 17.72 & 17.88  &18.32& 22.11 & 22.15 &22.53&  21.61 \\
			$15^-$ & 19.11&18.69& 19.69 & 23.31 & 22.82&25.18&26.54  &25.53&28.05& 26.45 & 22.07 & 21.90 &22.20& 27.55 & 26.97 &27.92&  27.30 \\
			$17^-$ & 23.11& 23.36& 23.03 & 28.27 & 28.67 &29.93&32.13  &32.13 & 33.67&32.43 & 26.71 &26.98  &26.14& 33.42 & 33.90 &33.50&  33.51 \\
			$19^-$ & 27.35 & 27.02& 26.39& 33.51 &32.97 &34.77&38.10  &37.21  &39.44& 38.91 &       &       && 39.63 & 39.68 &39.23&  40.24 \\
			$21^-$ &    &  & &38.99  &39.69 &39.68 &44.41  &44.76 &45.34& 45.87 &       &       &  &     &       & &      \\
			$23^-$ &    &   & &44.67&44.34 &44.63&51.03  &50.4  &51.32& 53.32 &       &       & &       &       & &       \\
			$25^-$ &   &  & &50.55  &51.86 &49.63& 57.93 &58.15 &57.38& 61.24 &       &       &&       &       &&        \\
			$27^-$ &   &  & &56.60 & 56.80 &54.66&65.08 &64.98 &63.49& 69.64 &       &       & &     &       &&       \\
			&  & &  & & & &  & & &  & & & & & & &\\
			\hline
	\end{tabular}}\label{Tab2}
\end{table*}
\par Of course, as we have already mentioned, the parameter $\sigma$ serves as a valuable metric for comparing the performance of the AQOA-S and AQOA-D models. Analyzing the data in the table, we note that both models yield relatively low values of $\sigma$ for ${ }^{222} \mathrm{Ra}$, indicating satisfactory fits to the experimental spectra. However, it becomes evident that the AQOA-S model surpasses the AQOA-D model, achieving a superior result with a lower value of $0.604$ compared to $0.917$ for AQOA-D. Similarly, for ${ }^{224} \mathrm{Ra}$ and ${ }^{226} \mathrm{Ra}$, the AQOA-S model consistently demonstrates superior performance, as evidenced by its lower $\sigma_{S}$ values of $0.801$ and $1.232$, respectively, compared to the larger values of $1.351$ and $1.360$ obtained by AQOA-D. These outcomes strongly suggest that the AQOA-S model offers a more accurate fit to the experimental data for these specific nuclei. Additionally, in the case of ${ }^{224} \mathrm{Th}$, the AQOA-S model maintains its advantage over AQOA-D, with a $\sigma_{S}$ value of $0.691$ compared to AQOA-D's $0.843$. However, it is noteworthy that for ${ }^{226} \mathrm{Th}$, the AQOA-D model surprisingly outperforms AQOA-S, delivering better results. Overall, these findings underscore the heightened accuracy and reliability of the AQOA-S model in reproducing the experimental spectra, as quantified by the quality measure $\sigma_{S}$.
%-----------------------------------------------------
%-----------------------------------------------------
\par After the completion of the fitting procedure, our focus shifts to determining the electromagnetic transitions E2, E1, and E3 in ${ }^{224} \mathrm{Ra}$. To evaluate the matrix elements for E2 transitions, we take into account parameters such as $b$, $F2$ (associated with the scale factor $t_2$), and $a_{s}$ (associated with the sextic potential). The value of b is derived by solving a quadratic equation, as described in equation (35) of Ref.\cite{b6}, using the ratios of experimental matrix elements. In the case of ${ }^{224} \mathrm{Ra}$, the average ratio leads to a value of $b=1.950$. The parameters F2 and $a_{s}$ are obtained through rms fitting, resulting in $F2=172.36$ and $a^{-1/4}_{s}= 1.24$. Moving on to E3 transitions, the value of $F3$ is determined by evaluating the ratio of E3 matrix elements to E2 matrix elements. For ${ }^{224} \mathrm{Ra}$, this ratio yields $F3=1123.68$ (or $(t_3/t_2)\sqrt{B_2/B_3}=6.519$). Finally, for E1 transitions, the quantity  $F1$ is treated as a global constant in accordance with Ref. \cite{b6}, and determined through rms fitting, resulting in $F1=5.37\times 10^{-3}$.
\par In Table \ref{Tab3}, we present a comprehensive comparison of matrix element values for electric transitions in ${ }^{224} \mathrm{Ra}$. The data includes measurements from experimental sources, expressed in units of $e$ fm, $e$ fm$^2$, $e$ fm$^3$ for E1, E2, and E3 transitions, respectively, and were obtained from \cite{b46}. Additionally, theoretical predictions derived from the AQOA-D model, sourced from Ref.\cite{b6}, are included for comparative purposes with both the experimental data and our AQOA-S model results.
\begin{table}
	\caption{Matrix element values of electric transitions in ${ }^{224} \mathrm{Ra}$. The experimental data, in units of $e$ fm, $e$ fm$^2$, $e$ fm$^3$ for E1, E2, and E3, respectively, have been taken from Ref. \cite{b46}. Theoretical predictions for the AQOA-D model, obtained from Ref. \cite{b6}, are used for comparison with both the experimental data and our AQOA-S model.}\label{Tab3}. 
	$
	\begin{array}{lccc}
		\hline\hline \text { Matrix element } & \text { Experiment} & \text {AQOA-D}& \text {AQOA-S} \\
		\hline\left\langle 0^{+}\|E 2\| 2^{+}\right\rangle & 199 \pm 3 & 196 & 193.06\\
		\left\langle 2^{+}\|E 2\| 4^{+}\right\rangle & 315 \pm 6 & 323 & 315.884 \\
		\left\langle 4^{+}\|E 2\| 6^{+}\right\rangle & 405 \pm 15 & 426&418.427\\
		\left\langle 6^{+}\|E 2\| 8^{+}\right\rangle & 500 \pm 60 & 525 &500.417\\
		\left\langle 1^{-}\|E 2\| 3^{-}\right\rangle & 230 \pm 11 & 236&230.399 \\
		\left\langle 3^{-}\|E 2\| 5^{-}\right\rangle & 410 \pm 60 & 334 &330.718\\
		\left\langle 0^{+}\|E 2\| 2^{+}_{\small{\tilde{\beta}}}\right\rangle & 23 \pm 4 & 36 & 43.1\\
		\left\langle 0^{+}\|E 3\| 3^{-}\right\rangle & 940 \pm 30 & 1006 &1003.69\\
		\left\langle 2^{+}\|E 3\| 1^{-}\right\rangle & 1370 \pm 140 & 1137 &1096.05\\
		\left\langle 2^{+}\|E 3\| 3^{-}\right\rangle & <4000 & 1176 &1156.86\\
		\left\langle 2^{+}\|E 3\| 5^{-}\right\rangle & 1410 \pm 190 & 1594 &1551.5\\
		\left\langle 0^{+}\|E 1\| 1^{-}\right\rangle & <0.018 & 0.013 &0.011\\
		\left\langle 2^{+}\|E 1\| 1^{-}\right\rangle & <0.03 & 0.018 &0.016\\
		\left\langle 2^{+}\|E 1\| 3^{-}\right\rangle & 0.026 \pm 0.005 & 0.023 &0.021\\
		\left\langle 4^{+}\|E 1\| 5^{-}\right\rangle & 0.030 \pm 0.010 & 0.032 &0.030\\
		\left\langle 6^{+}\|E 1\| 7^{-}\right\rangle & <0.10 & 0.042 &0.037\\
		\hline\hline
	\end{array}
	$
\end{table}
Upon examining the data, it becomes apparent that the AQOA-S model generally yields matrix elements that are closer to the experimental values compared to the AQOA-D model, except for a few specific matrix elements. For instance, in the case of the matrix element $\left\langle 0^{+}|E 2| 2^{+}\right\rangle$, the experimental value is $199\pm3$ $e$ fm$^2$, whereas the AQOA-S prediction of $196$ $e$ fm$^2$ is much lower to it than the AQOA-D prediction of $193.06$ $e$ fm$^2$. Similarly, for other matrix elements like $\left\langle 2^{+}|E 2| 4^{+}\right\rangle$ and $\left\langle 4^{+}|E 2| 6^{+}\right\rangle$, the AQOA-S model provides values ($315.884$ $e$ fm$^2$ and $418.427$ $e$ fm$^2$)  that exhibit better agreement with the experimental measurements ($315\pm6$  $e$ fm$^2$ and $405\pm15$ $e$ fm$^2$) compared to the AQOA-D model($323$ $e$ fm$^2$ and $426$  $e$ fm$^2$).
\par  
In the scenario of negative parity, the matrix element $\left\langle 1^{-}|E2| 3^{-}\right\rangle$ exhibits an experimental value of $230\pm11$  $e$ fm$^2$. Predictions by AQOA-D and AQOA-S yield values of $236$  $e$ fm$^2$ and $230.399$ $e$ fm$^2$ respectively. AQOA-S offers the closest approximation, whereas AQOA-D predicts a notably lower value. Similarly, for the matrix element $\left\langle 3^{-}|E2| 5^{-}\right\rangle$, the experimental value is $410\pm60$ $e$ fm$^2$. Predictions by AQOA-D and AQOA-S amount to $334$ $e$ fm$^2$ and $330.718$ $e$ fm$^2$ respectively. AQOA-S offers the most accurate prediction, while AQOA-D predicts a considerably lower value.  For the next matrix element, denoted as $\left\langle 0^{+}|E 2| 2^{+}_{\small{\tilde{\beta}}}\right\rangle$, which corresponds to positive parity, the experimental measurement yields a value of $23\pm4$$e$ fm$^2$. The theoretical predictions from AQOA-D and AQOA-S are $36$$e$ fm$^2$ and $43.1$$e$ fm$^2$, respectively. It is evident that both theoretical models considerably overestimate the experimental value.
\par Regarding the electric octupole transition, both AQOA-D and AQOA-S exhibit an overestimation of the experimental value, with AQOA-D deviating more significantly. Finally, when it comes to electric dipole transitions, there are certain matrix elements for which experimental values are unavailable, but upper limits have been determined. Interestingly, both AQOA-D and AQOA-S make predictions that are below these established upper limits. However, it is noteworthy that AQOA-S demonstrates slightly better agreement in terms of prediction accuracy compared to AQOA-D.
\par Ultimately, the AQOA-S model reveals much better agreement with the experimental data for the considered matrix elements in this study.

\section{Conclusions}
\par In summary, we  have presented the AQOA-S model, an extension of the analytic quadrupole octupole axially symmetric model, which incorporates for the first time a sextic potential to study nuclei with quadrupole-octupole deformation. By utilizing the sextic potential parameterized as $a_1\tilde \beta^2+a_2\tilde \beta^4+a_3\tilde \beta^6$, we derived analytical expressions for energy spectra and transition rates (B(E1), B(E2), B(E3)).
\par The application of the AQOA-S model to isotopes such as $^{222-226}$Ra and $^{224,226}$Th revealed that the energy spectra are governed by two crucial parameters: $\phi_0$, representing the balance between octupole and quadrupole strain, and $\alpha$, which affects the shape and behavior of the spectra through the sextic potential.
\par Our findings showed a remarkable agreement between the AQOA-S model and the recently determined B(EL) transition rates of  $^{224}$Ra, surpassing the performance of models employing the Davidson potential. This highlights the efficacy of our model in accurately capturing the stable octupole deformation observed in $^{224}$Ra.
\par The successful application of the AQOA-S model not only enhances our understanding of quadrupole-octupole deformation but also opens up possibilities for investigating a broader range of nuclei with similar characteristics. However, an additional noteworthy point to highlight here is that the adiabatic approximation used for $\phi$ in this study, relying on two steep harmonic oscillators, indeed comes with specific limitations.  It is widely recognized that achieving an accurate representation of parity splitting, often characterized as the odd-even staggering \cite{b17} of energy levels in the ground state band and the negative parity band, requires the presence of a finite, angular momentum-dependent barrier between the two potential wells \cite{b16,b47,b48}. As a result, this limitation leads to less precise theoretical predictions, especially for the odd-even staggering as well as the low-lying negative parity states, such as $1^-$ and $3^-$.
%--------------------------------------------------------------------------------------------------------
To address the inherent limitations of our moel regarding  these theoretical predictions, we plan to undertake a thorough investigation in a forthcoming  work where  the utilization of methodologies introduced either by Minkov et al.\cite{b17b} and by Budaca et al. will be inspected \cite{b49}.
%--------------------------------------------------------------------------------------------------------
\par Future research can also further explore the predictive capabilities of the AQOA-S model and its applicability to other nuclides with octupole deformation. Additionally, experimental validation of the model's predictions in different isotopes would strengthen its credibility and broaden its scope of applications.
\par Finally, our study contributes to advancing the field of nuclear structure and offers a powerful tool for investigating the intricate interplay between quadrupole and octupole deformations in atomic nuclei.


\begin{thebibliography}{99}
  \bibitem{b1}A. Bohr and B. R. Mottelson, Nuclear Structure, Vol. II (Benjamin, New York, 1975).
  \bibitem{b2}S. G. Rohozi\'nski, Rep. Prog. Phys. {\bf 51}, 541 (1988). 
  \bibitem{b3}I. Ahmad and P. A. Butler, Annu. Rev. Nucl. Part. Sci. {\bf 43}, 71 (1993). 
  \bibitem{b4}P. A. Butler and W. Nazarewicz, Rev. Mod. Phys. {\bf 68}, 349 (1996). 
  \bibitem{b5}D. Bonatsos, D. Lenis, N. Minkov, D. Petrellis, and P. Yotov, Phys. Rev. C {\bf 71}, 064309 (2005). 
  \bibitem{b6}D. Bonatsos, A. Martinou, N. Minkov, S. Karampagia, and D. Petrellis, Phys. Rev. C {\bf 91}, 054315 (2015).
  \bibitem{b7}J. Engel and F. Iachello, Phys. Rev. Lett. 54, 1126 (1985).
  \bibitem{b8}C. E. Alonso, J. M. Arias, A. Frank, H. M. Sofia, S. M. Lenzi, and A. Vitturi, Nucl. Phys. A586, 100 (1995).
  \bibitem{b9}A. A. Raduta and D. Ionescu, Phys. Rev. C 67, 044312 (2003), and references therein.
  \bibitem{b10}A. A. Raduta, D. Ionescu, I. Ursu, and A. Faessler, Nucl. Phys.A720, 43 (1996).
  \bibitem{b11}N. V. Zamfir and D. Kusnezov, Phys. Rev. C 63, 054306 (2001).
  \bibitem{b12}N. V. Zamfir and D. Kusnezov, Phys. Rev. C 67, 014305 (2003).
  \bibitem{b13}T. M. Shneidman, G. G. Adamian, N. V. Antonenko, R. Jolos, and W. Scheid, Phys. Lett. B 526, 322 (2002).
  \bibitem{b14}P. O. Lipas and J. P. Davidson, Nucl. Phys. 26, 80 (1961).
  \bibitem{b15}V. Y. Denisov and A. Dzyublik, Nucl. Phys. A589, 17 (1995).
  \bibitem{b16}R. V. Jolos and P. von Brentano, Phys. Rev. C 60, 064317 (1999).
  \bibitem{b17}N. Minkov, S. B. Drenska, P. P. Raychev, R. P. Roussev, and D. Bonatsos, Phys. Rev. C 63, 044305 (2001).
  \bibitem{b17a} R. V. Jolos , P. von Brentano  and F. Donau F, J. Phys. G: Nucl. Part. Phys. 19 L151 (1993).
  \bibitem{b17b} N. Minkov, P. Yotov, S. Drenska and W. Scheid, J. Phys. G: Nucl. Part. Phys. 32 497 (2006).
  \bibitem{b17c} R. V. Jolos, N. Minkov, and W. Scheid, Phys. Rev. C 72, 064312 (2005).
  \bibitem{b18}A. G. Ushveridze, Quasi-Exactly Solvable Models in Quantum  Mechanics (Institute of Physics Publishing, Bristol, 1994).
  \bibitem{b19}G. Lévai, J.M. Arias, Phys. Rev. C 69, 014304 (2004).
  \bibitem{b20}G. Lévai, J.M. Arias, Phys. Rev. C 81, 044304 (2010).
  \bibitem{b21}A.A. Raduta, P. Buganu, Phys. Rev. C 83, 034313 (2011).
  \bibitem{b22}A.A. Raduta, P. Buganu, J. Phys. G, Nucl. Part. Phys. 40, 025108 (2013).
  \bibitem{b23}R. Budaca, Phy. Let B 739,  56-61 (2014).
  \bibitem{b24}P. Buganu, R. Budaca, Phys. Rev. C 91, 014306 (2015).
  \bibitem{b25}P. Buganu, R. Budaca, J. Phys. G, Nucl. Part. Phys. 42, 105106 (2015).
  \bibitem{b26}R. Budaca, P. Buganu, M. Chabab, A. Lahbas and M. Oulne, Ann. Phys. (NY) 375, 65 (2016).
  \bibitem{b27}H. Sobhani, A. N. Ikot and H. Hassanabadi, Eur. Phys. J. Plus 132, 1-9 (2017).
  \bibitem{b28}R. Budaca, P. Buganu, and A. I. Budaca, Nucl. Phy. A 990, 137-148 (2019).
  \bibitem{b29}A. Lahbas, P. Buganu, R. Budaca, Mod. Phys. Lett. A 35,  2050085 (2020).
  \bibitem{b30}A. El Batoul, M. Oulne and I. Tagdamte, J. Phys. G 48, 085106 (2021).
  \bibitem{b31}G. Lévai and J. M. Arias, J. Phys. G 48, 085102 (2021).
  \bibitem{b32} M. Oulne and I. Tagdamte, Phys. Rev. C 106, 064313 (2022).
  \bibitem{b33}S. Baid, G. Lévai and J. M. Arias, J. Phys. G 50, 045104 (2023).
  \bibitem{b34}F. Iachello, Phys. Rev. Lett. 87, 052502 (2001). 
  \bibitem{b35} P. G. Bizzeti and A. M. Bizzeti-Sona, Phys. Rev. C 70, 064319 (2004).
  \bibitem{b36} P. G. Bizzeti and A. M. Bizzeti-Sona, Phys. Rev. C 77, 024320 (2008).
  \bibitem{b37}A. Ya. Dzyublik and V. Yu. Denisov, Yad. Fiz. 56, 30 (1993) ,[Phys. At. Nucl. 56, 303 (1993)].
  \bibitem{b38} V. Yu. Denisov and A. Ya. Dzyublik, Nucl. Phys. A 589, 17 (1995).
  \bibitem{b39}F. Iachello, Phys. Rev. Lett. 87, 052502 (2001).
  \bibitem{b40}F. Iachello, Phys. Rev. Lett. 85, 3580 (2000).
  \bibitem{b41}K. Heyde, Basic Ideas and Concepts in Nuclear Physics (IOP Publishing, Bristol, 1994).
  \bibitem{b42} %  Ra222-226 
  J. F. C. Cocks, P. A. Butler, K. J. Cann, P. T. Greenlees, G. D. Jones, 
  S. Asztalos, P. Bhattacharyya, R. Broda, R. M. Clark, M. A. Deleplanque, 
  R. M. Diamond, P. Fallon, B. Fornal, P. M. Jones, R. Julin, T. Lauritsen, 
  I. Y. Lee, A. O. Macchiavelli, R. W. MacLeod, J. F. Smith, F. S. Stephens, and C. T. Zhang,
  Observation of octupole structures in radon and radium isotopes and their contrasting behavior at high spin,  
  Phys. Rev. Lett. {\bf 78}, 2920 (1997). 
  \bibitem{b43} % Th230-234, Ra222-230
  J. F. C. Cocks, D. Hawcroft, N. Amzal, P. A. Butler, K. J. Cann, 
  P. T. Greenlees, G. D. Jones, S. Asztalos, R. M. Clark, 
  M. A. Deleplanque, R. M. Diamond, P. Fallon, I. Y. Lee, 
  A. O. Macchiavelli, R. W. MacLeod, F. S. Stephens, P. Jones, R. Julin, 
  R. Broda, B. Fornal, J. F. Smith, T. Lauritsen, P. Bhattacharyya, and C. T. Zhang,
  Spectroscopy of Rn, Ra and Th isotopes using multi-nucleon transfer reactions,  
  Nucl. Phys. A {\bf 645}, 61 (1999). 
  \bibitem{b44}
  A. Artna-Cohen, Nuclear Data Sheets for A = 224, Nucl. Data Sheets {\bf 80}, 227 (1997).  
  \bibitem{b45}
  Y. A. Akovali, Nuclear Data Sheets for A = 226, Nucl. Data Sheets {\bf 77}, 433 (1996).  
  \bibitem{b46}L. P. Gaffney et al., Studies of pear-shaped nuclei using accelerated radioactive beams, Nature (London) 497, 199
  (2013).
   \bibitem{b47}R.V. Jolos and P. von Brentano, Phys. Rev. C {\bf 49}  R2301 (1994). 
   \bibitem{b48}R. V. Jolos and P. von Brentano, Nucl. Phys. A {\bf  587} , 377 (1995).
   \bibitem{b49}R. Budaca, P. Buganu, and A. I. Budaca, Phys. Rev C 106, 014311 (2022).
\end{thebibliography}
\end{document}